%% file: main.tex
\documentclass[conference]{IEEEtran}
\def\showcomments{0}
\input{defs}
\ifCLASSINFOpdf
\else
\fi
\begin{document}
%
\title{Guarding Serverless Applications with SecLambda}

\makeatletter
\renewcommand\AB@affilsepx{, \protect\Affilfont}
\makeatother

\author[1]{Deepak Sirone Jegan}
\author[2]{Liang Wang}
\author[1]{Siddhant Bhagat} 
\author[3]{Thomas Ristenpart}
\author[1]{Michael Swift}
\affil[1]{UW-Madison}
\affil[2]{Princeton University}
\affil[3]{Cornell Tech}
\maketitle

\input{abstract}

\input{intro}


\input{motivation}

\input{threat}

\input{design}

\input{impl}

\input{app}

\input{eval}

\input{conclusion}




%
{\footnotesize \bibliographystyle{IEEEtran}
\bibliography{biblio}}

\end{document}

%% file: defs.tex
\usepackage{epsfig,endnotes}
\usepackage{etex}
\usepackage{setspace}
\usepackage{tikz}
\usepackage{framed,multicol}
\usetikzlibrary{shapes,arrows,fit,calc,positioning}
\usepackage{cite}
\usepackage{pgfplots}
\usetikzlibrary{matrix}
\usepgfplotslibrary{groupplots}
\usepackage{pgfplotstable}
\usepackage{multirow}
\usepackage{tabularx}
\usepackage{listings}
\usepackage{mdframed}
\usepackage[compact]{titlesec}
\usepackage[linesnumbered,ruled]{algorithm2e}
\usepackage{diagbox}
\usepackage{authblk}
\usepackage{algorithmic}
\usepackage{xspace}
\usepackage{array,graphicx}
\usepackage{booktabs}
\usepackage{pifont}
\usepackage{balance}
\usepackage{amssymb}
\usepackage[cmex10]{amsmath}
\usepackage{mathtools}
\usepackage{float}
\usepackage[export]{adjustbox}
\usepackage{paralist}
\usepackage{boxedminipage} 
\usepackage{color}
\usepackage[hyphens]{url}
\usepackage[breaklinks=true]{hyperref}
\hypersetup{colorlinks=true, breaklinks=true}
\usepackage{outline}
\usepackage{latexsym}
\usepackage{amsfonts}
\usepackage{footnote}     
\usepackage{wrapfig}      
\usepackage[]{caption}
\usepackage{subfig}
\usepackage{siunitx}
\usepackage{tikz}
\usetikzlibrary{patterns}
\usepgfplotslibrary{external} 
\usepgfplotslibrary[external] 
\usetikzlibrary{pgfplots.external} 
\usetikzlibrary[pgfplots.external] 

\ifnum\showcomments=1
    \newcommand{\fixme}[1]{{\textcolor{red}{[\textbf{FIXME}: #1]}}}
    \newcommand{\checkme}[1]{{\textcolor{orange}{[CHECKME: #1]}}}
    \newcommand{\liangw}[1]{{\textcolor{red}{[LW: #1]}}}
    \newcommand{\deepaks}[1]{{\textcolor{purple}{[DS: #1]}}}
    \newcommand{\draft}[1]{{\textcolor{darkgray}{[Draft: #1]}}}
    \newcommand{\tnote}[1]{{\textcolor{blue}{[TomR: #1]}}}
    \newcommand{\mnote}[1]{{\textcolor{cyan}{[Mike: #1]}}}

    \newcounter{mynote}[section]
    
    \newcommand{\thenote}{\thesection.\arabic{mynote}}
\else
    \newcommand{\fixme}[1]{}
    \newcommand{\checkme}[1]{}
    \newcommand{\thenote}[1]{}
    \newcommand{\liangw}[1]{}
    \newcommand{\draft}[1]{}
    \newcommand{\tnote}[1]{}
    \newcommand{\mnote}[1]{}
    \newcommand{\znote}[1]{}

\fi

\newcommand{\tabref}[1]{{Table~\ref{#1}}}
\newcommand{\figref}[1]{{Figure~\ref{#1}}}

\newcommand{\secref}[1]{{\S\ref{#1}}}

\newcommand{\bheading}[1]{{\vspace{4pt}\noindent\textbf{#1}}}

\makeatletter

\let\c@table\c@figure
\makeatother 

\newlength{\saveparindent}
\newlength{\saveparskip}
\newcounter{ctr}

\newenvironment{newitemize}{%
\begin{list}{\mbox{}\hspace{2pt}$\bullet$\hfill}{\labelwidth=10pt%
\labelsep=3pt \leftmargin=15pt \topsep=2pt%
\setlength{\listparindent}{\saveparindent}%
\setlength{\parsep}{\saveparskip}%
\setlength{\itemsep}{2pt} }}{\end{list}}

\newenvironment{newenum}{%
\begin{list}{{\rm (\arabic{ctr})}\hfill}{\usecounter{ctr} \labelwidth=17pt%
\labelsep=5pt \leftmargin=22pt \topsep=3pt%
\setlength{\listparindent}{\saveparindent}%
\setlength{\parsep}{\saveparskip}%
\setlength{\itemsep}{2pt} }}{\end{list}}

\newlength\figureheight
\newlength\figurewidth
\newlength{\arrow}
\newlength{\arrowl}
\makeatletter
\newcommand{\mydashleftrightarrow}[2][]{\ext@arrow 3359\leftrightarrowfill@@{#1}{#2}}
\def\rightarrowfill@@{\arrowfill@@\relax\relbar\rightarrow}
\def\leftarrowfill@@{\arrowfill@@\leftarrow\relbar\relax}
\def\leftrightarrowfill@@{\arrowfill@@\leftarrow\relbar\rightarrow}
\def\arrowfill@@#1#2#3#4{%
  $\m@th\thickmuskip0mu\medmuskip\thickmuskip\thinmuskip\thickmuskip
   \relax#4#1
   \xleaders\hbox{$#4#2$}\hfill
   #3$%
}
\makeatother

\newcolumntype{L}[1]{>{\raggedright\let\newline\\\arraybackslash\hspace{0pt}}m{#1}}
\newcolumntype{C}[1]{>{\centering\let\newline\\\arraybackslash\hspace{0pt}}m{#1}}
\newcolumntype{R}[1]{>{\raggedleft\let\newline\\\arraybackslash\hspace{0pt}}m{#1}}

\newcommand{\sysname}{SecLambda\xspace}
\newcommand{\helloapp}{\textsf{HelloRetail}\xspace}
\newcommand{\codeapp}{\textsf{CodePipline}\xspace}
\newcommand{\mapapp}{\textsf{MapReduce}\xspace}
\newcommand{\flow}[2]{{$#1$}$\rightarrow${$#2$}}

\definecolor{lightgray}{rgb}{.9,.9,.9}
\definecolor{darkgray}{rgb}{.4,.4,.4}
\definecolor{purple}{rgb}{0.65, 0.12, 0.82}

\lstdefinelanguage{JavaScript}{
  keywords={typeof, exports, new, true, req, send, status, false, catch, function, return, null, catch, switch, var, if, in, while, do, else, case, break},
  keywordstyle=\color{blue}\bfseries,
  ndkeywords={class, boolean, throw, implements, import, this},
  ndkeywordstyle=\color{darkgray}\bfseries,
  identifierstyle=\color{black},
  sensitive=false,
  comment=[l]{//},
  morecomment=[s]{/*}{*/},
  commentstyle=\color{purple}\ttfamily,
  stringstyle=\color{red}\ttfamily,
  morestring=[b]',
  morestring=[b]"
}

\lstdefinelanguage{Python}{
  keywords={def, replace, str, encrypt, get_token, send_action},
  keywordstyle=\color{blue}\bfseries,
  ndkeywords={class, boolean, throw, implements, import, this},
  ndkeywordstyle=\color{darkgray}\bfseries,
  identifierstyle=\color{black},
  sensitive=false,
  tabsize=4,
  comment=[l]{\#},
  morecomment=[s]{/*}{*/},
  commentstyle=\color{purple}\ttfamily,
  stringstyle=\color{red}\ttfamily,
  morestring=[b]',
  morestring=[b]"
}

%% file: abstract.tex
\begin{abstract}
As an emerging application paradigm, serverless computing attracts attention from more and more attackers. Unfortunately, security tools for conventional applications cannot be easily ported to serverless, and existing serverless security solutions are inadequate. In this paper, we present \emph{SecLambda}, an extensible security framework that leverages local function state and global application state to perform sophisticated security tasks to protect an application. We show how SecLambda can be used to achieve control flow integrity, credential protection, and rate limiting in serverless applications. We evaluate the performance overhead and security of SecLambda using realistic open-source applications, and our results suggest that SecLambda can mitigate several attacks while introducing relatively low performance overhead.    

\end{abstract}

%% file: intro.tex
\section{Introduction}

Serverless computing~(or function-as-a-service, FaaS) is an emerging application 
deployment architecture that completely hides server 
management from tenants. Serverless has a new programming model: 
an application is decomposed into small components, called \emph{functions}, each of which 
is a small application dedicated to specific tasks that runs in a dedicated \emph{function instance}~(a 
container or another kind of sandbox) with restricted resources such as CPU time and memory. 
A function instance, unlike a virtual machine~(VM) in traditional infrastructure-as-a-service
~(IaaS) platforms, will be launched only when there are requests for the function to process 
and is paused immediately after handling one request.
Tenants are charged on a 
per-invocation basis, without paying for unused and idle resources. 
Serverless has been used as a general programming model for a variety of
applications~\cite{jonas2017occupy,yan2016building,fouladi2017encoding}.

With the growing adoption of serverless, security challenges also arise. Like conventional 
web applications, vulnerabilities in the functions or third-party 
libraries being used can be exploited by attackers to subvert the control flow 
and data flow of applications to steal sensitive data and perform stealthy operations, 
as demonstrated in \cite{attack1, attack2, cryptomining}. Existing security tools for web applications 
have been ported to serverless, such as vulnerability scanning tools and 
log-based anomaly detection; but, they are only useful for detecting known vulnerabilities, 
or are not for real-time attack detection. We would like to complement those
approaches with real-time detection and prevention mechanisms.
Such tools~\cite{functionshield, intrinsic, vandium, epsagon} 
usually implement simplified information flow control by running functions in a sandbox and letting tenants 
specify and control the resources a function can or cannot access. However, 
such simple blacklist-based policies cannot detect manipulation of  
legitimate information flows, e.g., out-of-order or redundant flows. Besides, 
they only focus on securing each individual function and ignore the distributed nature of serverless. 
Because they lack visibility into the entire application, 
these tools fail to detect attacks that leverage incorrect function execution order~(i.e., invalid application execution paths) 
to subvert application logic~(See~\secref{sec:motivation}).
Overall, existing security tools are not sufficient for serverless applications. 


After surveying open-source serverless applications, we identify several common 
design patterns that could be leveraged to improve serverless security.  
In serverless, a function is \emph{stateless} to facilitate scaling 
up a large number of copies of the function for handling a high volume of requests. 
Hence, tenants need to externalize the data produced by the function 
to other services for late use to avoid data loss. Such externalization behaviors 
can be monitored and used for anomaly detection. The potentially complex communications 
between functions and third-party services could make understanding the 
expected behaviors of a serverless application much harder. However, as 
an application is decomposed into dedicated-task functions, it is possible 
for us to individually model each function, which usually has relatively simple logic, 
and construct a global view of the application. Besides, decomposition 
of an application makes it easier to enforce customized policies for different components, 
overall providing an opportunity for more flexible and efficient security
monitors.

Inspired by the insights from our survey, we design a novel serverless security framework that we call \emph{SecLambda}. 
In \sysname, a function runs in 
a modified container runtime environment \emph{runsec}
that intercepts certain function calls~(e.g., HTTP requests and I/O operations) 
and passes current function state   
to a \emph{guard} module. The guard 
executes a set of \emph{security functions}~(a piece of code dedicated to specific security tasks) 
to perform security tasks based on function state, and returns back the expected action that will be 
enforced by the runtime. A \emph{controller} centralizes management of security 
functions and policies, coordinates and collects function states from guards, 
and uses the global state to facilitate the guard to perform sophisticated security tasks.

The container runtime in \sysname 
is built atop the gVisor container runtime~\cite{gvisor}. We improve gVisor 
to support customized user-defined security policy that enables one to~(1) manipulate system calls 
based on high-level information such as network request payload and URL, and~(2) allow   
only specified system libraries in the guest OS to make system calls. Our improvements 
not only reduce the attack surface of gVisor-based containerized applications, but 
also facilitates the flexible control of application behaviors. The runsec runtime 
can serve as a building block for future gvisor-based security applications, and we will 
release its code publicly soon.

\sysname enables one to develop various security functions 
that can perform new security tasks as well 
as conventional security tasks~(e.g., SQL Injection detection). 
To show the flexibility of \sysname, we exercise \sysname to develop 
two novel security functions and one function for conventional security task: 
~(1) The \emph{flow tracking} function  
can assist a tenant in automatically modeling the expected interactions  
among different components within an application, and detect and block unexpected 
interactions, without cooperations from third-part services, 
to prevent sabotage of the control flow of an application. 
~(2) The \emph{credential management} function decouples credential data 
from function code, preventing an attacker's access to it
upon the compromise of credential-needing functions. 
~(3) The \emph{rate limiting} function restricts the number of instances of  
a given function running in a certain time period and the rates of given 
API requests generated from a function to mitigate DoS attacks against the application and prevent 
the application from being used for DoS.

We implement a prototype of \sysname, and evaluate the 
performance overheads \sysname and the security of flow tracking function 
with various applications and workloads. 
Our analysis suggests \sysname introduces relatively small overhead while preventing  
attacks that cannot be handled by existing serverless security tools. 





%% file: motivation.tex
\section{Motivation and Challenges}~\label{sec:motivation}
In this section, we introduce different types of attacks against serverless, 
the limitations of existing serverless security tools, common design patterns 
of serverless applications that inspire our system design, and  
the challenges in developing efficient security mechanisms in serverless.

\subsection{Attacks against serverless}
Serverless has drawn attention from attackers and researchers, 
and several attacks, which mostly tamper with control flow, have been proposed. 
We define two types of control flow for serverless application: for an application 
that consists of multiple functions, the order in which its functions are executed 
while handling user requests is defined as \emph{application control flow}; and for 
a function, the order in which requests are sent from in its executions 
is \emph{function control flow}. 
We illustrate an example application in \figref{fig:example} and three 
types of control flow related attacks that motivate our system design: 

\begin{figure}[t]
\begin{center}
\includegraphics[width=0.99\linewidth]{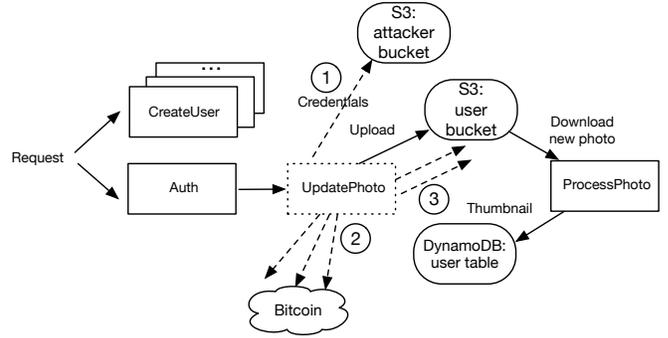}
\end{center}
\caption{An example of serverless application. In the application, the \texttt{UpdatesPhoto} 
function will uploads the photo sent by an authenticated user to S3, which will then trigger 
the \texttt{ProcessPhoto} function to generate the thumbnail of the newly-uploaded photo 
and store the thumbnail into DynamoDB. The rectangle with dotted lines represent compromised 
functions, and dotted lines are flow injected by the attacker.}
\label{fig:example}
\end{figure}

(1)~The attacker hijacks both application and function control flow. Examples 
are \textbf{data exfiltration attacks} and \textbf{crypto-mining attacks}~\cite{attack1, cryptomining}. 
In data exfiltration attacks, the attacker exploits vulnerable functions, 
i.e., functions with vulnerabilities in the libraries or code, to steal sensitive 
data stored in the function code or environment variables. For example, in \figref{fig:example} part~(1), 
the attacker extracts the AWS credentials from the \texttt{UpdatesPhoto} 
function and sends them to her own S3 bucket. In crypto-mining attacks, 
as in \figref{fig:example}~(2), the attacker performs her activity stealthily at the cost of the tenant.
A concrete attack proposed by PureSec~\cite{cryptomining} shows that the attacker can 
turn one single vulnerable function into a virtual crypto-mining farm without being noticed by the tenant. 

(2)~The attacker tampers with function control flow but does not affect application control flow 
by exploiting legitimate execution paths of the application.  
In \figref{fig:example}~(3), the attacker performs DoS attacks against \texttt{ProcessPhoto}
by uploading a large amount of data from \texttt{UpdatesPhoto} to the S3 bucket. 

(3)~The attacker manipulates function execution order~(i.e., application control flow) 
to subvert application logic, while function control flow is not affected. For example, 
one directly invokes \texttt{UpdatePhoto} without being authenticated by \texttt{AUTH}. 

\subsection{A survey on serverless applications}
To guide our development of defenses against the aforementioned attacks, we survey open-source 
serverless applications to understand their common design patterns. 
We search ``serverless'' and ``lambda'' on GitHub to find the most popular repositories 
for AWS-based serverless applications that are written in Python or Node.js,  
remove toy applications, and manually inspect the remaining 50 
applications~(40 are provided by AWS).~\footnote{We conducted this survey in 2019.}
The applications consist of a total of 86 functions.~(Note that a single-function application may still have a large codebase.)
Below we highlight three patterns that we believe should be considered when developing 
security-enhancing mechanisms for serverless applications.

\begin{newitemize}
\item{\emph{No local storage.}}
In general, no examined applications store data~(e.g., 
intermediate processing results) on local disks due to 
the \emph{stateless} nature of serverless. Serverless providers do not guarantee 
that requests can always be handled by the same function. So storing stateful 
data on local disk faces the risks of data loss and application crash. The common 
practice for passing data across requests is to store data on some storage service~(s) 
and retrieve it later.    

\item{\emph{No direct interactions between functions.}}
There are usually no direct information flows between functions, except in 
two applications. In most cases, applications rely on other services to transfer 
control from one function to another function. To invoke function $B$, function $A$ 
may send events to some service that can forward the events to $B$ 
or the service will automatically trigger $B$ upon certain operations of $A$. 
Transferring data across functions is the same as across requests, relying on storage services. 
In fact, in AWS Lambda, the size of function input is limited to 6\,MB~(for synchronous requests) 
or 256\,KB~(for asynchronous requests), which may not be sufficient for some applications. 
To safely transfer arbitrary data between functions or requests, one should use storage services. 

\item{\emph{Input-dependent functions are common.}}
We examined the number of HTTPS connections made by a function~(no function issues HTTP or socket connections), 
and found for about 40\% of the functions the number of requests and their destinations depend on input parameters. 
More specifically,~(1) 52 of the functions do not make any HTTPS requests~(e.g., HTTP header rewriter 
and image processing functions) or hard-code the requests in the code.~(2) For 23 functions, 
we can estimate the maximum number of requests they could make~(e.g., by examining how many times a function calls given APIs), 
and the exact set of request destinations. The exact number of requests during execution, 
however, may depend on function input. Many serverless applications adopt this pattern: 
they only set up one serverless function as an entry point, and perform different operations 
based on user input. In fact, this is the fastest way to port conventional  
web applications~(e.g., Django-based web applications) to serverless.~(3) For 11 functions, the number 
of requests or request URLs may vary according to user input. For example, one logging application takes 
a set of events as input, processes each event, and stores them into different S3 buckets;   
a weather application fetches weather information for all the cities mentioned 
in the input from an API service, who uses different URLs for different cities,  
and sends back an aggregated result. 
\end{newitemize}

Overall, HTTPS requests from a function indicate either a data transfer or 
a control transfer. By monitoring HTTPS requests within an application, we can 
therefore monitor the control flow of the application. For most of the functions, 
the destinations of their requests are known, which means 
we are able to model their behaviors. For some functions, we can only know 
the requested URLs at runtime. However, one useful observation is that the URLs 
requested by such functions have a fixed pattern, mostly in the form of 
``common prefix + variable''. Knowing these URL patterns is also helpful for 
modeling function control flow. Though there could be other patterns, we only consider 
this one in our project.

\subsection{Challenges}
To prevent the previously discussed attacks, we need to accurately track the control  
flow of an application.  
Existing information flow protection mechanisms usually monitor system calls of interest 
by modifying OS or system libraries~\cite{zeldovich2008securing, pasquier2017camflow}. 
Such mechanisms often assume an endpoint can be identified by IP and port, which 
is not sufficient for serverless functions.  
In serverless, 
a function is associated with dynamically-assigned and ever-changing IPs, and a 
service might have the same IP and port as other services~(e.g., 
one can redirect requests to different services on the same host based on 
the \texttt{Server Name Indication} field). 
So the URL is necessary for endpoint identification, but it is 
difficult to directly extract high-level information such as URL and HTTP header from 
the low-level information seen by such mechanisms. 
Considering the limited resources in function instances, we need to wisely  
choose the granularity of monitoring to reduce monitoring overhead and 
capture more meaningful information at the same time. 

Furthermore, since serverless functions may be input-dependent, i.e., 
number of requests and endpoint URLs vary based on input parameters, we should be able 
to capture and model such behaviors. 

Another challenge is tracking control flow across heterogeneous infrastructures. 
As mentioned before, a serverless application often relies on certain 
functionalities provided by third-party services, such as hosted 
databases. It is unrealistic to assume that all the services 
are willing to upgrade their infrastructures to support new security mechanisms. 

Finally, for ease of development, the new security mechanism should be application-independent 
and is transparent to applications, i.e., applications do not need to be modified.

\subsection{Limitations of existing serverless security tools}
Many services and tools have been developed for improving serverless security. 
One such service is AWS X-Ray, which is a tracing service provided by AWS for monitoring requests 
sent to/from different components of an AWS-based serverless application~\cite{xray}. However, 
one must modify their applications~(i.e., no application transparency) to use X-Ray libraries, 
which could be a considerable amount of extra work for large applications. Besides, 
X-Ray can only track requests created by several high-level libraries~(e.g., \emph{httplib} in Python), 
so it is easy to circumvent its monitoring via a lower-level library. 
For example, in Python, X-Ray is not aware of any requests created via the \emph{socket} module. 
AWS WAF is a web application firewall for protecting AWS-based web applications~\cite{waf}. 
Unfortunately, WAF is used for inspecting the requests from outside of AWS. 
Communications between the functions in a serverless application do not go through WAF.

Another class of security tools for serverless run a function in a sandbox, 
and monitor the network connections~(or file operations) made by the function. Several 
tools~\cite{intrinsic, epsagon, functionshield, vandium} block connections based on the 
policies --- usually just IP/URL whitelists --- provided by the tenant. 
Such tools fail to handle application control flow hijacking due to the lack of global 
visibility of the entire application, and 
can only prevent function control flow hijacking to some extent. As in \figref{fig:example}, 
S3 is a legitimate destination for \texttt{UploadPhoto}; therefore, a compromised 
\texttt{UploadPhoto} can make an arbitrary number of requests to S3 when 
using whitelist-based access control, though only one should be allowed. Trapeze~\cite{trapeze}
and Valve~\cite{valve} enable fine-grained control of information flow in serverless 
applications and provide stronger security guarantees. Unfortunately, they cannot 
handle input-dependent functions and uncooperative third-party services. We will discuss 
more on Trapeze and Valve in \secref{sec:related}.










%% file: threat.tex
\section{Threat Model}

We consider three major parties in our threat model: a target application, 
an attacker, and third-party services. 
The target application is deployed on a serverless computing platform by a trustworthy 
tenant~(the application owner). An application might consist of 
multiple functions. Any services/applications/functions, other than the 
functions in the target application, are considered to be \emph{third parties}, 
including services from the same 
cloud provider and the services set up by the same tenant outside the 
serverless platform. 

We treat a third-party service as a blackbox that takes input from some 
sources and outputs results to some destinations. 
Both the input sources and output 
destinations of the third-party service must be within the target application. 
The exact functions that generate an input and receive the corresponding output 
might be different, though. For example, as in \figref{fig:example}, \texttt{UpdatePhoto} uploads a 
picture~(input) to S3, and S3 will generate an ``upload'' event~(output) to trigger 
\texttt{ProcessPhoto} to process the picture. 

The application may store sensitive authentication data such as 
encryption keys or access tokens within the function's code and, therefore,
each function instance. 

We assume that the attacker can compromise at least one 
function of the application, leveraging bugs in the function code, 
vulnerable libraries used in the functions, or inappropriate configurations. 
The attacker is able to perform any type of control flow related attacks discussed in~~\secref{sec:motivation}.
We further assume that all the operations the attacker can perform must be done via the functions.

We assume that the platform itself is secure. By that we mean that the adversary
cannot compromise host VMs, serverless runtime, or 
third-party services. Nor can they manipulate network traffic.

%% file: design.tex
\section{Design of \sysname}\label{sec:design}

\begin{figure}[t]
\begin{center}
\includegraphics[width=0.99\linewidth]{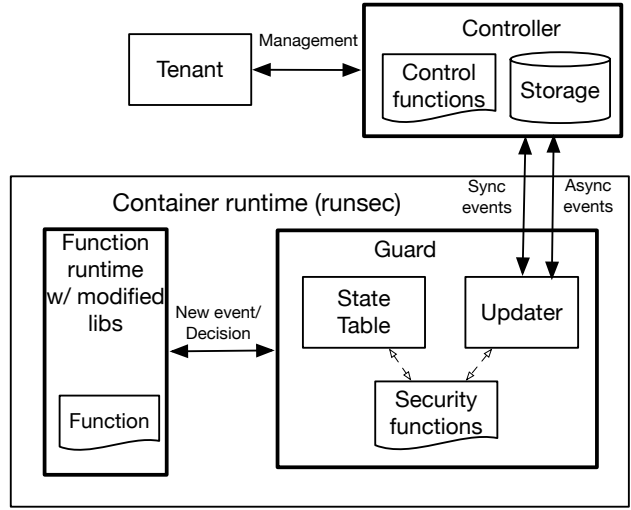}
\end{center}
\caption{An overview of \sysname architecture.}
\label{fig:guard_arch}
\end{figure}

Motivated by the above discussions, we design a novel system 
for securing serverless application that we call \emph{SecLambda}, which 
allows tenants to run customized security tasks to protect their applications 
and supports security policy of flexible type.    
As in \figref{fig:guard_arch}, 
\sysname consists of three basic components: function runtime, guard, 
and controller.   


\bheading{Secure runtime.} 
The \sysname secure runtime consists of a modified container runtime \emph{runsec} and 
a function runtime with instrumented system libraries. 
The container and function runtimes are instrumented to generate and report 
 important \emph{events} to an event processing module that we call it~\emph{guard}. 
 An event indicates the current operation performed by the function, 
 including (1) send/receive messages over SSL/TLS connections, (2) 
send/receive messages over sockets, and~(3) create/modify/delete files. 
Network connection and file metadata~(destination IP, port, filename, etc.), 
request payload, and file content, can also be sent in the event.  
The container runtime blocks an operation of interest until 
it receives a decision from the guard, and enforces the decision. 
The decision is either \textsf{ALLOW} or \textsf{DENY}, plus the additional action the runtime should take, 
such as modifying the operation parameters. We use gVisor~\cite{gvisor}, a lightweight container runtime, 
to build our secure runtime, and will elaborate on its implementation and security in \secref{sec:impl}.

\bheading{Controller.} 
The controller provides a centralized interface for 
a tenant to manage and distribute \emph{security functions}, policies, and configurations. 
A security function is 
a piece of code dedicated to a specific security task such as access control and rate limiting. 
It also serves as a centralized logger and collects function states from all function instances  
to maintain the global state of the application. 
The global state will be used as the input for \emph{controller-side control functions}, which 
facilitate the runtime to make decisions on certain function operations.

\bheading{Guard.} 
The guard is a module of runsec, which executes a set of  
security functions created by the tenant to process the received events 
based on the tenant's security policies.  
Based on the outputs from the security functions, the guard determines the action 
to be taken for the operation associated with an event, and returns 
the decision to the runtime. Some security functions may require knowledge from 
the controller. In this case, the guard will communicate with the controller via the \emph{updater}, 
which is responsible for the communication between the guard and the controller, and 
forward the decision from the controller back to the runtime. 
The guard also forwards any received events to the controller asynchronously via the updater.

The updater waits for updates from the controller, and updates the other components accordingly. 
An update can be a new policy or function configuration.   
A state table maintains the aggregated  
function state, e.g., the number of HTTPS connections issued by the function.

\bheading{Guard registration.}
When a function instance is launched, runsec will generate generate a random ID for the 
function instance, and sends  
the instance ID and the function name to the controller. The controller creates 
a 16-byte unique guard ID based on instance function and function name, and sends the guard ID 
back to the guard. In future communications~(sending events, reconnection, etc.), 
the guard simply carries the guard ID in the events for authentication. 
The controller maintains a map between guard ID, instance ID, function name, and application name. 
The mapping between function name and application name is provided by 
the tenant. The function name is usually  
stored in the environment variables, as common practice for serverless platforms~\cite{wang2018peeking}.   

All the data needed by the security functions,  
such as security policies, user credentials, and other configurations,  
will be downloaded from the controller during instance startup.

\bheading{Event format.}
The event exchanged between the guard and the controller has a fixed-length, 22-byte header 
consisting of the following fields: guard\_id~(16 bytes), body\_length~(4 bytes), 
function\_type~(2 bytes), and event\_type~(2 bytes). The function\_type field indicates 
the security function the event is associated, and event\_type indicates 
the purpose of the event. The body part of an event stores the data~(e.g., security policy) 
being sent, the format of which is specified by the associated security function. 
The event sent between the runtime and the guard will not contain the guard\_id field 
in the header.

%% file: impl.tex
\section{\sysname implementation}\label{sec:impl}

\subsection{Secure runtime for serverless functions}
\sysname container runtime \emph{runsec} is built atop gVisor~\cite{gvisor}, a lightweight 
container runtime developed by Google. The gVisor runtime provides an emulation 
of the Linux kernel over which a containerized application is traced either with a \emph{ptrace} 
system call or by running the container in a minimal virtual machine using $kvm$. 
Briefly, whenever the containerized application makes a system call, it is intercepted 
by the gVisor runtime which then handles the call as Linux would. \sysname leverages 
this capability of gVisor to interpose on system calls made by the function to 
enforce various policies.  

We modify the gVisor runtime to forward events 
and the relevant call parameters to the guard for selected system calls,
and block the system calls until getting decisions from the guard.   
Specifically, we instrument the \texttt{SendMsg} and \texttt{Write} system calls for our evaluation.

For HTTPS requests, the data is encrypted so the runtime cannot see 
the plaintext messages. In this case, the runtime restricts the function to use 
an instrumented version of the OpenSSL library~(i.e., \texttt{libSSL}) that 
passes the plaintext data to gVisor before it is encrypted. Considering 
that serverless platforms become increasingly flexible, i.e., supporting C/C++ based 
functions, the function may be able to load external libraries to bypass the call 
monitoring. To address this issue, we add a new hypercall \texttt{Validate} 
in the gVisor runtime to perform \emph{code page validation}.
Each socket descriptor has to be passed to 
the \texttt{Validate} hypercall 
before it can be used for connecting and sending operations. 
The code page from where the \texttt{Validate} hypercall is invoked is 
checked to see if it is within the code pages of the mapped \texttt{libSSL} library. 
This technique can be applied to ensure that all the permitted operations come from 
instrumented or audited system libraries, and not from externally injected code. 

The system calls in Linux are not atomic. Checking system call arguments for policy adherence could lead to race 
conditions where the arguments to a system call made by a thread 
are altered by another thread, after a check is performed on them. For example, a buffer used for sending network data can be modified by another thread in the same application after the check passes for the thread calling the system call but before the send is actually performed.
Such a Time-of-Check-to-Time-of-Use~(TOCTOU) bug was demonstrated in the Janus  
system call interposition framework~\cite{wagner1999janus, garfinkel2003traps}. We assume that each serverless 
function is single-threaded and hence to the best of our knowledge we believe 
that our implementation does not suffer from the TOCTOU bug. 
Note that currently 
\sysname does not support multi-threaded serverless functions, and we leave handling 
multi-threated functions as future work. 

\subsection{Guard and controller}
The guard is implemented as a module in the gVisor runtime. 
the gVisor routine~(i.e., \texttt{runsc-sandbox}) 
that intercepts system calls will call the relevant function provided 
by the guard to pass events. The guard launches the updater in a goroutine during 
instance startup, and communicates with 
the controller using \emph{zeromq}~\cite{zeromq}. The guard maintains two 
long-lived \texttt{zeromq} connections with the controller, one for sending 
synchronous events that require the decisions from the controller, and one 
for sending all events asynchronously to the controller for logging purposes. 

The controller is implemented in C/C++ and the other components are implemented in GO, 
totaling about 1\,K lines of C/C++ code, and 2.5\,K lines of GO code. 
We are still in the process of developing a full set APIs that can be used for developing 
security functions and the management interface.

\subsection{A prototype of \sysname for AWS}\label{sec:prototype:aws}
Our real implantation of \sysname is built on gVisor, and we use it to run a performance benchmark 
to understand real performance in a local testbed~(see \secref{sec:eval:local}). We separately 
implemented a prototype of \sysname of limited functionality that 
can run in AWS Lambda, and use it to measure the impact of AWS network latencies, 
which may be different than the local testbed, on \sysname performance. 
This prototype also allows us to perform security evaluation on real 
AWS-based serverless applications. In this prototype, we rewrite the guard module 
as a standalone program in C/C++, and the controller remains unchanged. We will explain 
how this prototype will be used in \secref{sec:eval:flow}.

%% file: app.tex
\section{Security Function: Flow tracking}\label{sec:flowtracking}

All of the previously discussed attacks alter the control flow, 
either application control flow or function control flow, of an application. 
To enforce Control Flow Integrity~(CFI) in serverless application, we 
borrow the ideas from model-based IDS and the original CFI 
technique~\cite{wagner2001intrusion, giffin2002detecting, giffin2004efficient, cfi}. 
For application control flow,  
we treat an application as a single program and each of its functions as a basic block in the original CFI~\cite{cfi}, 
and add checks when a function receives requests and returns responses. 
For function control flow, we built a model of the acceptable request sequences 
of a function, and monitor the requests sent by the function.   


Several challenges exist for achieving CFI in serverless: understanding control flow of an 
application and its functions, constructing good policies~(i.e., control flow graphs), 
and enforcing the policies efficiently. Static analysis may not work well for Python 
and Node.js based serverless applications, 
and manual analysis is not cost-effective. We nevertheless overcome these challenges 
and develop a semi-automated mechanism that we call it flow tracking function.

\bheading{Overview.}
%
We define a \emph{flow} as 
one message exchange between a function and another endpoint~(a service or a function).
We assume all the functions support \sysname while services do not.  
Let $f$ be a function, $m_{in}$ be the message $f$ received from the 
standard function entry point, and $m_{out}$ be the message $f$ returns via the standard 
exit point. Note that $f$ can only receive one $m_{in}$ and send one $m_{out}$ in an execution. 
$m_{req}$ is a message sent from $f$ to another endpoint $s$, 
and $m_{resp}$ is the response from $s$. $m_{req}$ and $m_{resp}$ are sent over the 
same connection. We currently do not know any serverless platforms 
that support direct access to functions via their IP addresses, so $m_{in}$ 
and $m_{resp}$ correspond with the only two ways for passing messages to functions.
The flow tracking security function can automatically generate \emph{flow graphs} 
for a target application and enforce function and application 
control flow integrity given a flow graph. There are two types of flow graphs: 
global graph and local graph. 
A global graph is a directed graph, where a node represents a function, and an edge 
from node $A$ to node $B$~(denoted by \flow{A}{B}) represents that function $A$ sends 
messages~(directly or indirectly) to function $B$ from connections initialized by $A$. 
Each function is associated with a local graph, wherein a node presents 
an operation performed~(e.g., an HTTPs request) by the function, 
and an edge points to the next expected 
operation. For simplicity of discussion, we only consider network requests 
when building the local graph.

The flow graph generated by our security function is a variant of 
flow-sensitive and context-sensitive nondeterministic finite automaton~(NFA) that 
models the expected flow sequences of a function or valid execution paths of an application. 
To construct such NFAs, we trade off space for model accuracy by duplicating states and 
removing cycles from NFAs as in the IAM model~\cite{gopalakrishna2005efficient}. 
Next, we discuss how to generate these graphs and enforce them. 

\bheading{Handling user-input via URL replacement.}
To identify the URLs that may be constructed based on user input, 
we group all the URLs extracted from the traces associated with a function
by their longest common prefix~(LCP) with the following restriction: 
given a set of URLs $U = \{u_1,...,u_n\}$, two URLs $u_i$ and $u_j$ are grouped 
only if their longest common prefix $LCP(u_i, u_j)$ is longer than $LCP(u_i, u_k)$ 
and $LCP(u_j, u_k)$ for any $u_k \in U$ ($k \neq i, k \neq j$). 
Then, if the number of unique URLs in a group is more than a threshold $t_{lcp}$, we replace 
the URL in a flow with the LCP of the group the URL belongs to, and append 
a ``$*$'' to the LCP to distinguish it from regular URLs. We call the resulting URLs the \emph{LCP URLs}. 
An example is that three URLs \emph{a.com}, \emph{a.com/test/x}, and \emph{a.com/test/y} 
will produce two LCP URLs~(two groups) \emph{a.com} and \emph{a.com/test/*}. 
We consider that the URLs in the same group are more related with each other. 
See \secref{sec:eval:security} for more discussions.

\bheading{Generating flow graphs.}
For simple applications, the flow graphs can be created manually by the tenant. To 
reduce human errors involved during graph creation, \sysname also provides a tool 
to create the flow graphs dynamically based on the collected traces. A trace 
is the sequence of flows generated by given functions in 
one \emph{application} execution, and records important information such as timestamps, function names, 
destination URLs, and HTTP operations~(\texttt{GET}, \texttt{POST}, etc.). To collect traces, 
the tenant should prepare and run test cases that can cover all the possible 
execution paths of an application. The test cases should send requests to 
the \emph{entry functions}, i.e., the functions that will accept requests 
from users, of the application to trigger executions of the application.  
Then, \sysname leverages the method proposed in Synoptic~\cite{beschastnikh2011leveraging}, 
which is originally designed for building loop-free NFAs from syscall traces, 
to generate flow graphs. 

In the local graph, each node 
is a $\langle$URL, HTTP operation$\rangle$ tuple that indicates that the function 
sends a message to an endpoint whose address is \emph{URL}. The local graph 
has an entry node indicating the endpoint from whom $f$ receives $m_{in}$, 
and an exit node indicating sending out $m_{out}$. 
In a \emph{completed} function execution, $f$ must follow one path from the entry node to 
the exit node. In the global graph, 
each node represents a function. We say that if the execution of a function $\bar{f}$ depends 
on the output of another function $f$ but no explicit messages are exchanged between $f$ and $\bar{f}$, 
there is an \emph{implicit flow} from $f$ to $\bar{f}$. One such example is that $f$ uploads 
a file to AWS S3, which generates a message to trigger $\bar{f}$ to process the uploaded 
file. We assume that a function  
depends on the function invoked immediately before it, and use the global graph  
to capture implicit flows.

\begin{figure}[t]
\begin{center}
\includegraphics[width=0.8\linewidth]{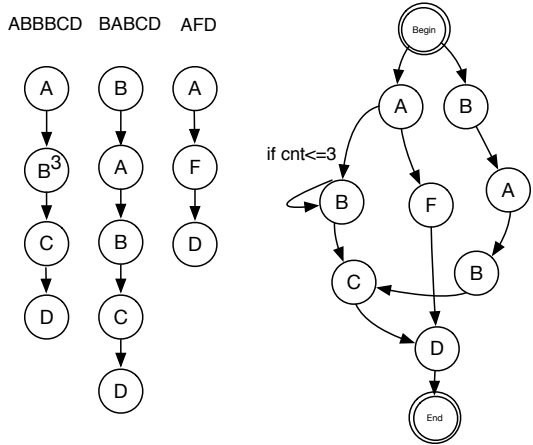}
\end{center}
\caption{An example of the generated local flow graph for three traces ABBBCD, BABCD, and AFD.}
\label{fig:nfa_example}
\end{figure}

To have more strict policies, we maintain \emph{loop counters} for each graph to 
restrict the number of repetitions of given flows or flow sequences.  For a trace of length $l$, 
we look for consecutive repeated subsequence(s) of length 1, ..., $\lfloor{l/2}\rfloor$. 
Such subsequence indicates the function sending a set of requests repeatedly. 
We treat such subsequences as a single flow, or \emph{grouped flow}, 
and use a loop counter to count the 
repetition of the subsequence.  We only need to maintain counters 
for the nodes whose counter is greater than one. An example is shown in \figref{fig:nfa_example}.

\bheading{Policy enforcement for the local graph.}
The guard checks the current function state against the generated local flow graph.
Upon sending a message, the guard proceeds to the successor node~(state) 
of the current node if the flow matches the successor node, and otherwise block the flow. 
If a node is associated with an LCP URL, the prefix of the destination URL in a 
legitimate flow should match the LCP. When it is at a node representing a grouped flow, 
the guard expects to see all the subsequent flows from the function match the flows 
associated with the grouped flow in order. 

\bheading{Policy enforcement for the global graph.}
We use \emph{tags}, which perform the same role as the labels in the original CFI, 
to carry function identity information. 
A tag is in the format of $\langle$function name, guard ID, request ID$\rangle$, 
while the request ID is just a random 16-byte string assigned to each request, 
generated by the entry function of the application. The tags are carried 
in the HTTP headers so guards can remove or add tags without changing messages, 
and services that do not support \sysname will simply ignore this header 
field and process the messages as~normal.

All guards run the following protocol:

(1)~If $m_{in}$ comes from another function 
in the application, the guard extracts the tag $\bar{t}$ from $m_{in}$ and checks the function name in $\bar{t}$ to see if 
the source of $m_{in}$ is legitimate, as specified in its local flow graph. Otherwise, the guard checks with the controller, 
who maintains the global flow graph and the global state of the application,   
to see if its preceding function $\bar{f}$ in the global flow graph~($\bar{f}{\rightarrow}s{\rightarrow}f$) 
is at a legitimate state, i.e., whether $\bar{f}$ has sent messages to $s$. 
Recall that the guard will forward received events asynchronously to 
the controller so the controller is aware of the states of all functions~(See \secref{sec:design}). 
The guard then gets the tag of $\bar{f}$ from the controller and saves it as $\bar{t}$.  

(2)~The guard reuses the request ID in $\bar{t}$ to generate its tag $t$.

(3)~The guard adds $t$ to $m_{req}$, and tries to remove the tag $t'$ from $m_{resp}$ ($t'$ 
may not exist if $m_{resp}$ comes from third-party services). 
The guard will check if $t$ and $t'$~(if it exists) are the same to make sure $m_{resp}$ is sent 
for it. 

(4)~The guard adds $t$ to $m_{out}$. If $m_{out}$ is sent to a function, 
the request ID in $t$ will be propagated to that function.

Note that serverless applications by nature do not satisfy the Non-Executable Data~(NXD) requirement 
in CFI~\cite{cfi}. Both Python and JavaScript enable users to execute a piece of 
string as code, which are the features that are commonly exploited by attackers. 
Therefore, the global graph itself can not prevent attackers from 
injecting arbitrary flows in the application. For instance, a compromised 
$f$ can initialize more than the expected number of \flow{f}{s}. 
But by using local flow graphs, such attacks can be detected.

\bheading{Concurrent requests.}
During trace collection, we only send one request at a time to the application; 
however, the application may receive multiple user requests simultaneously in practice, 
and multiple instances of the same function can be running concurrently. 
Though the enforcement of local flow graphs in this case will not be affected, 
it will be difficult for the controller to identify whose tag should be given to whom. 
In this case, the controller maintains a pointer for each application execution to record 
the current state of the execution individually. And, it maintains 
a list for each node in the global flow graph, 
and adds the corresponding function executions to the list in order based on their start times 
~(i.e., $m_{in}$ receiving times). The function executions at the same position 
of the lists for $f$ and $\bar{f}$ will be paired together. 
Here we do not try to accurately identify the causality of function executions, 
but only to ensure all the application executions and 
the implicit flows from any given function are legitimate. If it  
finds there are still executions left in $f$' list while $\bar{f}$'s list has no available executions for pairing, 
the controller will raise the alarm since this indicates violating execution paths.     
We will discuss more in \secref{sec:eval:security}.

\section{More security functions}

In this section we will introduce the security functions for credential protection 
and rate limiting. By using them and flow tracking function together, an application 
can achieve better security. 
 
\subsection{Credential management}

AWS suggests tenants use AWS's Key Management Service~(KMS) and Secrets Manager 
Service~(SMS) to protect their credentials~\cite{kms, sms}. Unfortunately,  
the credentials stored in these services need to be decrypted in the function before being 
used for API calls, so they are still in plaintext in memory. 
A malicious module may be able to access these credentials in the caller function~(e.g., 
using the \texttt{inspect} method in Python). 

We use \sysname to secure API credentials by decoupling the API credentials 
from the functions and storing them in the guard. 
When a message that carries credentials 
is sent out, the guard will rewrite the message to add the credentials 
at specified locations. If a response contains sensitive data, the guard 
can extract and store the sensitive data for later use, 
and obfuscate the sensitive data that will be accessed by the functions.  

Though services use standard authentication or authorization protocols, the exact message formats 
in different APIs can vary. So the credential management function is application-independent 
but not service-independent; that is, one needs to understand the semantics 
of those APIs, which is usually publicly accessible via 
API documents or API libraries, to develop the security functions.
  
\figref{fig:example:cognito} shows an example credential management function 
for AWS Cognito, a user management and authentication service provided by AWS. 
In a Cognito-integrated application, all the user 
information~(password, phone number, etc.) will be stored and managed by Cognito. 
When a user logs into the application, the user credentials will be sent to Cognito 
for authentication, and several 
tokens will be sent back for the user to perform follow-up operations. One important 
token is called an \emph{access token}, which allows the user to access and modify her 
own information. In the example, we aim at protecting the credentials of a special 
user --- the administrator, because with her credentials an attacker 
can access any user information.

\begin{figure}[t]
\center
\begin{minipage}{0.48\textwidth}
		
		\lstset{frame=single,xleftmargin=1.0ex, basicstyle=\tiny}
		\begin{lstlisting}
		client = boto3.client('cognito-idp')
		resp = client.initiate_auth(
			...# public variables
				AuthParameters={
			'USERNAME': 'placeholder_a',
			'PASSWORD': 'placeholder_b'
			})

		token = resp['AuthenticationResult']['AccessToken']
		print token # random string

		resp = client.list_devices(AccessToken=token)

		\end{lstlisting}
		
\end{minipage}
\begin{minipage}{0.48\textwidth}
		\lstset{frame=single,xleftmargin=0.9ex, basicstyle=\tiny}
		\begin{lstlisting}
	...
	admin_name = "test"
	admin_pwd = "123456"
	auth_url = "https://cognito-idp.us-east-2.amazonaws.com/" 
	auth_op = "InitiateAuth"
	token_extractor = "..." # a reg exp for extracing access token from string
	auth_msg_tmpl = "..., AuthParameters={'USERNAME': admin_name, 
	'PASSWORD': amdin_pwd}" # auto generated
	other_reqs = [{url:a, op:b}, ...]
	# the above are from user policy 

	# store the HTTP request in the "http" object
	If event is "send" and http.url is auth_url and http.op is auth_op:
		new_msg = gen_msg(auth_msg_tmpl) # generate new message
		wait_resp = True
		auth_session = http.session_id
		# ask runtime to send the new message
		send_action("allow/update_msg", new_msg) 
	If event is "recv" and wait_resp = True and http.session_id = auth_session:
		access_token = get_token(http.payload, token_extractor) # extract token
		#ask runtime to replace the token with a random string
		send_action("allow/replace", access_token, encrypt(access_token)) 
		wait_resp = False
	If http.url/http.op pair mathes one in the other_reqs:
		if event is "send":
	 		send_action("allow/replace", encrypt(access_token), access_token)
	 	if event is "recv":
	 		send_action("allow/replace", access_token, encrypt(access_token))

		\end{lstlisting}
\end{minipage}

\caption{An example of user login function~(\textbf{upper}) and 
the credential management function~(\textbf{lower}) for AWS Cognito APIs. }
\label{fig:example:cognito}
\vspace{-0.5cm}
\end{figure}

There are two common ways for carrying credentials in 
the messages:~(1) in the HTTP \texttt{Authorization} header, and~(2) in the 
message body. Cognito uses the second method. 
In the login function in \figref{fig:example:cognito}, the tenant uses placeholders 
in place of the username and password, but stores the real username and password 
in the guard. Before the authentication message is 
sent out, the guard will check if the message is for Cognito authentication by 
looking at the destination URL and the other information in the HTTP message. 
The URL information is sufficient for determining whether the message is sent to 
the right endpoint, while the specific API operation performed can be revealed from 
the URL, other header fields~(e.g., in AWS it might be in the \texttt{X-Amz-Target} header), 
or~(rarely) the body. If the message is legitimate, the guard 
will reconstruct a message that contains the correct 
credentials, and let the runtime send the new message. 
When receiving the response for the message, which should contain the access token, 
the guard extracts the access token using regular expressions from the response and replaces the access 
token in the response with a random string, before the runtime returns the 
response to the function. From the view of the function, it only sees 
some dummy credentials and an invalid access token that can never be used anywhere else. 
For the follow-up API requests, the function will use the invalid access token. The guard 
will replace the invalid token with the actual token in the requests, 
and replace the actual token with the invalid token in the responses.

To reconstruct the authentication message, the security function generates 
a message based on the template of the authentication message, and fills 
the message with the credentials specified in the user policy and other information from 
the originally sent message. If credentials are carried in the \texttt{Authorization} header,  
we only need to delete the original \texttt{Authorization} header and add a new one, 
which usually has a simple format. For example, it is just \emph{Authorization: Bearer access\_token} 
in some AWS and Google APIs.  To reduce the manual labor of template creation, 
we develop a tool for automatically generating the message template of a given API request 
from OpenAPI specification~files~\cite{openapi}.

Our example security function can be reused for other APIs that have a similar 
workflow as the AWS Cognito APIs. Our framework allows the tenants to specify 
the regular expressions for extracting the API operation and the sensitive data 
from a message in the policy, rather than hardcoding them in 
the security function. 

\bheading{Message integrity check.}
Some APIs may perform integrity check or sign the entire message (e.g. the newest AWS 
S3 APIs). To handle these cases, the guard needs to produce the digests or signatures 
using required algorithms, which requires more effort at understanding APIs. However, 
we believe the offered security benefits are worth the extra~effort.

\subsection{Rate Limiting}
We consider two types of rate limiting tasks:~(1) limiting the fan-out~(i.e., the number 
of running instances) of a given function, 
and~(2) limiting requests to a given API. For~(1), 
since every guard will register with the controller during startup, it is easy for 
the controller to maintain a per-function counter to 
track the number of launched instances for a given function. However, 
the provider might launch new instances to execute functions, and discard the 
existing instances. The controller may see a large number of new instances over 
time but only a portion of them are active. To handle this case, the controller 
will check if the instances are active by periodically~(i.e., every second) sending   
heartbeat messages, and decrease the corresponding counters for those 
instances that are no longer active.  

Using the controller 
as a centralized state store, we can restrict the frequency of requests sent to 
a given URL for a given operation from a function of interest. A rate limiting policy  
is a list of entries $<$function name, target URL, HTTP operation, rate$>$. 
Once a tenant deploys the rate limiting functions and policies on 
the guard and the controller, the controller will keep track of the events from functions, and 
calculate the rate of a given type of flow. If the rate of the flow exceeds 
the allowed rate, the controller will broadcast a \textsf{STOP} event to all 
the guards that are responsible for the functions generating the flow. 
Then, the guards will simply drop the flow, whose type is specified in the \textsf{STOP} 
event, until receiving a \textsf{RESUME} event from the controller.

%% file: eval.tex
\section{Evaluation}\label{sec:eval}

In our evaluation, we evaluate the security and performance overhead of \sysname. 
We focus on the flow tracking function in the evaluation. 
We start with security evaluation of three open-source serverless applications with our AWS-based 
prototype introduced in \secref{sec:prototype:aws}, and conduct a security analysis. 
For performance evaluation, we run our real implementation of \sysname (i.e., gVisor-based) 
in a local testbed to measure the runtime overhead, and use the AWS-based prototype 
to understand the impact of AWS network latencies on \sysname performance.  

\begin{figure*}[t]
\begin{center}
\includegraphics[width=0.90\linewidth]{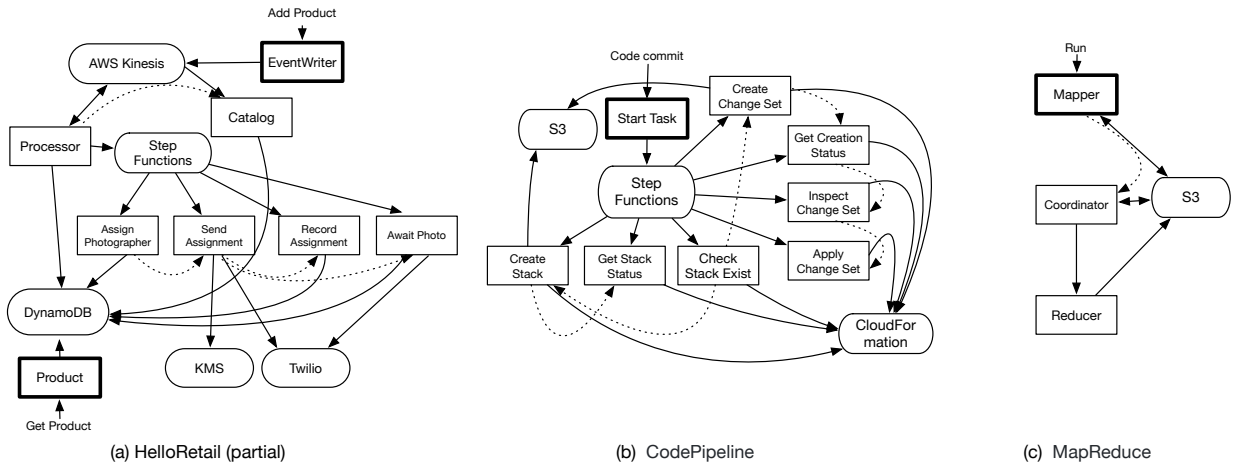}
\end{center}
\caption{Flow graphs of the target applications. Rectangles, rounded rectangles
and dotted lines represent function, third-party services, and implicit flows, respectively. 
The entry functions that accept user requests are highlighted. }
\label{fig:app_flow}
\end{figure*}

\subsection{Efficiency of flow tracking function}\label{sec:eval:flow}

\begin{table}[t]
\center
\begin{tabular}{|l|r|r|r|r|}
\hline
 & \multicolumn{1}{l|}{\textbf{LOC}} & \multicolumn{1}{l|}{\textbf{\#Func}} & \multicolumn{1}{l|}{\textbf{\#Lib}} & \multicolumn{1}{l|}{\textbf{Language}} \\ \hline \hline
\helloapp & 5,127 & 12 & 158 & Node.js \\ \hline
\codeapp & 2068 & 9 & 112 & Node.js \\ \hline
\mapapp & 7,47 & 3 & 1 & Python \\ \hline
\end{tabular}
\caption{An overview of the lines of code, number of functions and third-party modules, and 
the languages of the target applications.}
\label{tab:app_overview}
\end{table}

We choose for our evaluation three open-source Lambda-based serverless applications: \helloapp~\cite{hello}, 
\codeapp~\cite{code}, and \mapapp~\cite{map}. \helloapp, a retail platform developed by Nordstrom, 
is the most sophisticated open-source serverless application we have known. 
\codeapp is an application from AWS for automatically updating the deployment 
script of software after its source code or configuration has been modified. 
\mapapp is a serverless-based MapReduce framework. 
The three applications demonstrate the use of different 
features of serverless: \helloapp takes advantage of the event sourcing mechanism 
and AWS Step Functions to pass messages among functions and invoke function automatically,  
\codeapp purely leverages AWS Step Functions to automatically execute functions in order, 
and \mapapp relies on the auto-scaling feature to run functions in parallel. 
\tabref{tab:app_overview} and \figref{fig:app_flow} show an overview of the three applications 
and flow graphs. 

Note that Trapeze~\cite{trapeze} and Valve~\cite{valve} offer modified versions of \helloapp 
that can run locally via replacing certain cloud services with local counterparts. 
However, these modified versions do not emulate important services such as AWS Step 
Functions and KMS, which are core components of the application. We believe 
running applications in a native serverless environment can 
produce more realistic and accurate flows. 

\bheading{Setup of AWS-based \sysname prototype.}
In AWS Lambda, we cannot change the provided function runtime. 
As a workaround, we modify the Python and Node.js modules that 
are used by our target applications for sending requests to send events to the guard, and instrument the 
functions to use the modified modules. As for evaluation, 
we only need to modify the ssl module~(\texttt{ssl.py}) in Python, 
and the aws-sdk modules~(only \texttt{aws-sdk/libs/http/node.js}) 
in Node.js. The guard is compiled as a binary, and every function 
is instrumented to launch the guard asynchronously in a 
background process before processing events. Without modifying the infrastructures of AWS, 
\sysname cannot achieve full functionality. Nevertheless, AWS provides a 
realistic environment for more accurately evaluating the accuracy of 
auto-generated flow graphs. 

The controller ran on a m5.large VM in the AWS us-east-1 region, and all the experiments were 
conducted in the same region unless otherwise stated.

\bheading{Flow graph generation.}
We started with writing a minimum set of test cases that cover all 
the execution paths of the target applications.
Based on their documentation and state machines~(provided by Step Functions), 
we created 4, 3, and 1 test cases for 
\helloapp, \codeapp, and \mapapp, respectively. Several functions are triggered only  
when there are errors during request processing. In one round of test, 
we run all the test cases once with random user input. In \helloapp, user 
requests are for creating products, to register photographers, and to query products, so 
we generate random product/photographer information and query strings. \codeapp 
will be triggered automatically when there are changes to the monitored repository, 
and we add 1--5 random files~(1\,KB) in the monitored repository. For \mapapp, 
the job is fixed as word count, and user requests specify the dataset and number 
of files to be processed. We randomly choose one dataset from the \emph{text/tiny/rankings} and 
\emph{text/tiny/uservisits} datasets 
in the Big Data Benchmark from AMPLab~\cite{bigdata}, and process 1--9 files in a 
request~(the datasets only have 9 files). All the other settings remain default. 

We obey the limits~(request rate, input size, etc.) of all the APIs when constructing the requests, 
because violating these limits will not generate any new paths, but may cause 
Step Functions to be stuck for minutes, which unnecessarily prolongs the experiment time.   

We ran the test for 1,000 rounds. After analyzing the collected traces, we find  
a clear gap between expected flows and real flows from a function. 
Only two functions are associated with consistent flows in any test, 
while neither function actually makes any outgoing connections. For 
the rest of the functions, we collected varied numbers of unique traces. Note that 
the extra flows do not necessarily affect the global flow graph~(i.e., 
no new execution paths).    

We identify three unexpected causes for the varied flows:

\begin{newenum}
\item \emph{API library implementation}: 
API libraries may add randomness to endpoints' URLs. 
For instance, the AWS API for sending data to AWS Codepipeline~(used by 
the \texttt{CreateChangeSet} function in \codeapp) will automatically append a short,  
random string to the URL of AWS Codepipeline to distinguish different requests.  

\item \emph{Service configuration}: Server-side configuration such as HTTP redirection 
will introduce extra flows that cannot be inferred from code. 

\item \emph{Network failures or other unknown service behaviors}: In some cases, the 
function will retry a request if the request failed due to unstable network condition 
or received duplicated messages from services.

\end{newenum}

As expected, flows may vary according to user input: In \mapapp, the \texttt{Mapper} function 
will download data from S3 multiple times depending on the size of user data. Similarly, 
the \texttt{Coordinator} function will create different numbers of reducers based on user data. 
Besides, the exact URLs for the files in S3 will also vary.  

The final local graphs contain only 1--9 nodes~(excluding the begin and end nodes). 
The median size~(i.e., number of nodes) of the local graphs for \helloapp, \codeapp, 
and \mapapp are 3, 2, and 7, respectively. The global graphs generated 
are consistent with the state machines in Step Functions or the expected 
graph got from manual inspection. Maintaining these graphs in memory incurs negligible 
memory overhead.

\begin{figure}[t]
\begin{center}
\input{figure/graph_error.tex}
\end{center}
\caption{Errors when using the traces collected from different numbers of rounds  
for building flow graphs.}
\label{fig:flowerror}
\end{figure}
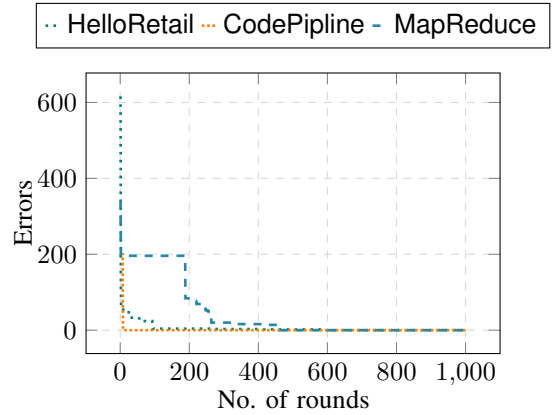

\bheading{Graph accuracy.}
Intuitively, more rounds of tests will produce more accurate flow graphs.
Based on the collected traces, we further estimate when using the traces from the first $n$ rounds 
to build flow graphs, the number of rounds the guards will block flows and cause function failures. 
As shown in \figref{fig:flowerror}, the traces from one round of test can cover 
the requests in about 40\%, 80\%, and 65\% of rounds, and no errors when using 583, 7, and 463 rounds of traces    
in \helloapp, \codeapp, and \mapapp respectively.

\subsection{Security analysis}\label{sec:eval:security}

The flow graphs generated by our flow tracking function represent relatively 
tight security policies. Even with a slight difference between the expected URL and its URL, 
a flow will be considered as invalid and be blocked. The generated flow graphs 
do not contain cycles or loops, and any given path in a graph represents 
a legitimate order of flows or function executions. Therefore, the 
flow tracking function can prevent an attacker from generating arbitrary flows from the 
compromised function, and reduces the risk of data leakage via restricting the ways 
for the attacker to externalize data. The loop counter associated with a node indicates 
how many times the flow can occur in a normal function execution. 
If it occurs more than the expected times, the flow will be blocked to prevent potential attacks.   

The attacker may try to bypass some functions~(e.g., authentication) to 
invoke a target function directly. For instance, the attacker invokes the \texttt{CreateChangeSet} 
function in \codeapp directly to modify the source code of a software. Though 
the resulting flows are legitimate, the flow tracking function may still block such attempts: 
if the requests come from invalid sources or 
do not carry correct tags, or the preceding function of the target function 
is not at a correct state, the execution will 
be blocked. Basically, we do not allow an application execution to begin from 
the middle of an execution path. The attacker may try to exploit the race condition 
between function executions to therefore perform a type of flow injection attacks. For example, 
the attacker leverages some vulnerabilities in Step Functions to execute  
\texttt{CreateChangeSet} just before a normal execution of \texttt{CreateChangeSet} begins. 
In this case, the controller will assume the attacker-launched execution is legitimate, 
but can still detect anomalies~(See \secref{sec:flowtracking} \emph{Concurrent requests}).        

As we discussed in \secref{sec:eval:flow}, static analysis 
will not be sufficient for understanding the flows generated under realistic workloads. 
Using dynamic analysis, security depends on the quality of test cases. 
The flow graphs will be more accurate with test cases that can cover more paths.
However, we may still encounter uncovered cases during normal application executions, 
which could cause execution failures. One possible option is to switch 
from fail closed to fail open, i.e., the system records suspicious flows 
rather than block~them.

The nodes with the LCP URLs 
allow the flows whose destination URLs share the same prefix, which may cause security issues. 
However, without them the resulting graphs could be too restrictive  
so that they cannot handle change in user input. 
For example in \mapapp, without using LCP URLs, the flows for 
fetching different files~(who have the same prefix and are in the format of 
``prefix0000'',...,``prefix000n'') will produce distinguish nodes and any future 
requests that are not for fetching these files will be blocked. We believe 
manual inspection of flow graphs is a reasonable way to solve this issue. 
The tenant can adjust the threshold being used to produce the optimal flow graphs 
and examine whether the nodes with the LCP URLs are appropriate.   


The flow tracking function cannot prevent attacks that exploit    
both legitimate function and application control flows, and data-related attacks~(e.g., 
modify the data sent by a compromised function). However, it is feasible to extend our 
framework and develop more sophisticated security functions to achieve 
finer-grained information flow control for serverless.

\subsection{Performance overhead of \sysname}\label{sec:eval:local}


\bheading{Local testbed setup.}
We set up a single control plane Kubernetes cluster of five nodes in CloudLab~\cite{cloudlab}, with each node running 
on a machine of an Intel Xeon 2.40GHz CPU and 64\,GB RAM.  
Each machine is connected to a star topology LAN network with a speed of 25\,Gbps. 
The \sysname controller runs on a separate identical node outside the LAN but on the same datacenter. 

\bheading{Runtime overhead.}
We create a Python-based  
test function to send 1000 HTTP identical GET request to a web server in the same cluster. 
We run the function in a container of $m$ memory. To emulate AWS Lambda, 
the CPU power assigned to the container is proportional to its memory based on 
the observation in \cite{wang2018peeking}. In the experiment, we adjust $m$ to 128, 256, 512, 1,024, and 2,048\,MB, 
and invoke the function with and without \sysname running. 

The guard is initialized with a local flow graph that has one path with one node, but 
the guard checks the single node 1000 times to emulate the worst-case policy lookup time 
of a 1000-node graph. When enforcing the local flow graph, the guard only needs to 
maintain one pointer that points to the current flow. The time of moving the pointer is negligible, 
so is the policy lookup time. The latency overhead of 1,000 policy lookup is 
less than \SI{100}{\micro\second} when $m$ = 128\,MB. We believe 1,000  
is a reasonably large size for a local flow graph, so 
for ordinarily applications the policy lookup overhead is sufficiently low, and 
only accounts for a negligible portion of the overhead introduced by \sysname.

\begin{figure}[t]
\begin{center}
\input{fig_lat.tex}
\end{center}
\caption{The average request completion time of 1,000 requests in \sysname and gVisor under different container 
memory sizes. Y axis is truncated at 400.}
\label{fig:completion}
\end{figure}
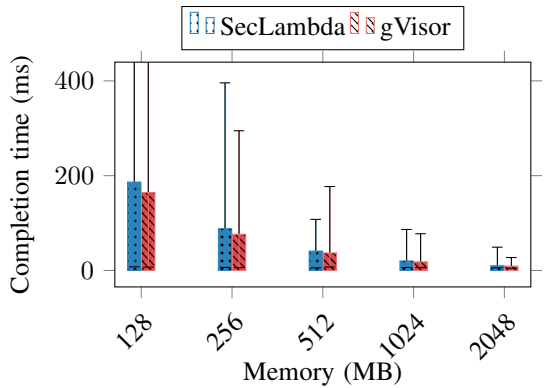

The difference of request completion times without~(i.e., using stock gVisor) 
and with \sysname is used to estimate the overhead introduced by runsec. 
\figref{fig:completion} shows the average request completion time under different settings. 
As expected, the runtime overhead increases as function memory decreases, because less CPU power is 
allocated. When compared with using stock gVisor, the average relative runtime overhead 
across all requests is 21.8\,ms~(13\%) when $m$ = 128\,MB, and only 0.7\,ms~(8\%) 
when $m$ = 2,048\,MB. Overall, the runtime overhead of \sysname is moderate.

\bheading{End-to-end latency of DeathStarBench applications.}
To understated the potential performance impact of \sysname on realistic applications, 
we further run the DeathStarBench benchmark suite~\cite{gan2019open} with \sysname in the local testbed, 
DeathStarBench provides a set of microservice applications, a workload generator \emph{wrk2} for generating predefined workloads for different 
applications, and a trace analysis tool that reports various performance metrics on each function as well as the entire application. 
We measure the end-to-end latencies of two DeathStarBench under various workloads: 
\begin{newitemize}
	\item social-network: The application implements a social network with unidirectional follow relationships between users. The implementation consists of loosely coupled long-running functions communicating with each other via Thrift RPC. 
	The built-in workloads for social-network are:
	\begin{itemize}
		\item compose-post (CP): The workload involves sending POST requests to an application entry function. The data is text only with the size varying within 1KB. The request is to post the payload text on a specific user's timeline.
		\item read-home-timeline (RHT): The workload sends GET requests to an application entry function. The size of the GET requests are within 100 bytes. 
		\item read-user-timeline (RUT): This workload is similar to the read-home-timeline workload, but the requests are sent to a different entry function.
	\end{itemize}
	\item media-microsvc: The application is a movie review service where users can sign up and post reviews about movies. Similar to social-network, media-microsvc also uses Thrift RPC to communicate between different functions.
	The only provided workload for media-microsvc is:
	\begin{itemize}
		\item compose-review (CR): The workload involves sending POST requests containing a movie review. The size of each request is within 500 bytes.
	\end{itemize}
\end{newitemize}
We show an overview of our target applications and workloads in~\tabref{tab:deathstar}.

The guard runs the flow tracking function. The policy generated for social-network 
has 253 nodes and 75 edges, while the policy for media-microsvc has a total of 210 nodes 
and 98 edges. We do not use the DeathStarBench applications to evaluate the efficiency 
of the flow tracking function because these applications do not exhibit typical 
serverless behaviors, which may cause \sysname fail to generate accurate policies. 
For instance, a long-running function in social-network keeps listening for requests 
from other functions after being launched, and can generate arbitrary numbers 
of flows without exiting. However, the generated 
flow graphs only allow explicit numbers of flows from the function to prevent DoS attacks, 
and thus may incorrectly block flows from the function. Nevertheless, 
DeathStarBench applications are developed based on realistic services, making 
them suitable for performance benchmarking.

\begin{table}[t]
\footnotesize
\centering
\renewcommand\tabcolsep{1.3pt}
\begin{tabular}{|l|l|r|r|r|r|r|r|}
\hline
\multicolumn{1}{|c|}{\textbf{App}} & \multicolumn{1}{c|}{\textbf{Workload}} & \multicolumn{1}{l|}{\textbf{\begin{tabular}[c]{@{}l@{}}LOC\\ (K)\end{tabular}}} & \multicolumn{1}{c|}{\textbf{\begin{tabular}[c]{@{}c@{}}\# Covered \\ Func\end{tabular}}} & \multicolumn{1}{l|}{\textbf{\begin{tabular}[c]{@{}l@{}}Max\\ Depth\end{tabular}}} & \multicolumn{1}{c|}{\textbf{\begin{tabular}[c]{@{}c@{}}SecLambda\\ Lat. (ms)\end{tabular}}} & \multicolumn{1}{c|}{\textbf{\begin{tabular}[c]{@{}c@{}}gVisor \\ Lat. (ms)\end{tabular}}} & \multicolumn{1}{c|}{\textbf{\begin{tabular}[c]{@{}c@{}}Overhead \\ (\%)\end{tabular}}} \\ \hline
\hline
\multirow{3}{*}{social} & CP & 5.7 & 11 & 4 & 26.92 & 25.86 & 4.07 \\ \cline{2-8} 
 & RHT & 0.9 & 2 & 2 & 8.63 & 8.38 & 2.89 \\ \cline{2-8} 
 & RUT & 1.2 & 2 & 2 & 8.74 & 8.38 & 4.20 \\ \hline
media & CR & 4.3 & 9 & 4 & 23.14 & 21.90 & 5.60 \\ \hline
\end{tabular}
\caption{An overview of DeathStarBench applications and workloads, and the average end-to-end latencies of the target applications 
under different workloads across 1,000 runs, with and without \sysname. ``Max Depth'' is the number of functions in the 
longest application execution path under a given workload.}
\label{tab:deathstar}
\end{table}

We ran each workload for 1,000 times and measured the average end-to-end latency across all runs. 
As in \tabref{tab:deathstar}, \sysname introduces relatively small latency overhead for different workloads, 
even though a workload may trigger the execution of 10 functions and tens of network requests between functions. 
The reason is that many functions are executed in parallel in the target applications. 
For real-world applications with parallel execution paths, 
the latency overhead introduced by \sysname may not be directly related to the number 
of operations their functions have performed. Therefore, \sysname could have little 
performance impact on real-world applications in practice.

\begin{table}[t]
\center
\renewcommand\tabcolsep{2pt}
\begin{tabular}{|l|r|r|r|r|}
\hline
 & \multicolumn{1}{c|}{\textbf{128}} & \multicolumn{1}{c|}{\textbf{512}} & \multicolumn{1}{c|}{\textbf{1024}} & \multicolumn{1}{c|}{\textbf{2048}} \\ \hline
 \hline
\textbf{1\,KB} & 65.6 (13.3) & 11.6 (6.1) & 6.5 (2.3) & 5.1 (0.7) \\ \hline
\textbf{10\,KB} & 69.8 (15.0) & 11.3 (5.5) & 7.4 (2.1) & 5.2 (0.6) \\ \hline
\textbf{1\,MB} & 945.2 (143.1) & 222.4 (11.9) & 100.6 (5.3) & 61.7 (2.0) \\ \hline
\end{tabular}
\caption{The average controller communication latencies and standard derivations~(in ms) across 
100 rounds under different message sizes and memory sizes.}
\label{tab:ctr_rtt}
\end{table}

\bheading{Controller communication overhead.}
Finally, we estimate the performance impact of running \sysname in AWS with our AWS-based 
prototype. Prior study shows AWS Lambda suffers from unpredictable network performance~\cite{wang2018peeking}, 
which may affect the round-trip latency between the guard and the controller. 
The round-trip latency can affect function running time, because the guard will communicate 
with the controller synchronously if the function 
receives messages from third-party services. 
We measured the round-trip latency between the guard and the controller 
under different message sizes and function memory sizes. We measured the latency
100 times for a given setting. The results are shown in \tabref{tab:ctr_rtt}. 
Considering that the messages exchanged between the guard and the controller are 
less than 1\,KB, and in one function execution the guard  
communicates with the controller synchronously at most once, the overhead 
introduced by guard-controller communications is around 6\,ms   
for functions of large memory sizes~(i.e., 1,024\,MB). Since Lambda measures 
runtime in 100\,ms blocks, the overhead only costs an iota of extra 
money~(\$0.000001667 per function). We also find that functions might also have 
unpredictable performance due to unstable network conditions and resource contention, 
especially for those functions that interact with third-party services. 
The running time of such functions in our target applications range 
from 128\,ms to 1,300\,ms. As such, the controller communication overhead is acceptable.

%% file: figure/graph_error.tex
\definecolor{color1}{rgb}{0.98, 0.81, 0.69}
\definecolor{color2}{rgb}{0.55, 0.71, 0.0}
\definecolor{color3}{rgb}{1.0, 0.6, 0.4}
\definecolor{color4}{HTML}{2B83BA}

\definecolor{bblue}{HTML}{4F81BD}
\definecolor{rred}{HTML}{D7191C}

\pgfplotsset{
compat=1.11,
legend image code/.code={
\draw[mark repeat=2,mark phase=2]
plot coordinates {
(0cm,0cm)
(0.15cm,0cm)        
(0.2cm,0cm)         
};%
}
}

\begin{tikzpicture}
\begin{axis}
[
y tick label style={
        /pgf/number format/.cd,
            fixed,
            fixed zerofill,
            precision=0,
        /tikz/.cd
    },
width = 0.8\columnwidth, 
height = .6\columnwidth,
ylabel = Errors,
xlabel = No. of rounds,
xlabel style={name=xlabel, yshift=0.5em},
ylabel style={name=ylabel, yshift=-0.5em},
grid = major,
grid style={dashed, gray!30},
legend style={at={(0.5,1.25)}, anchor=north, legend columns=-1},
cycle list name=exotic,
]

\addplot+[mark=none, line width=1pt, dotted] 
  table[x=no, y=hello, col sep=comma] {figure/data/graph_error.csv};
  \addplot+[mark=none,  line width=1pt, densely dotted] 
  table[x=no, y=code, col sep=comma] {figure/data/graph_error.csv};
  \addplot+[mark=none,  line width=1pt, dashed] 
  table[x=no, y=map, col sep=comma] {figure/data/graph_error.csv};
\legend{\helloapp, \codeapp, \mapapp}
\end{axis}
\end{tikzpicture}

%% file: fig_lat.tex

\definecolor{color1}{rgb}{0.98, 0.81, 0.69}
\definecolor{color2}{rgb}{0.55, 0.71, 0.0}
\definecolor{color3}{rgb}{1.0, 0.6, 0.4}
\definecolor{color4}{HTML}{2B83BA}

\definecolor{bblue}{HTML}{4F81BD}
\definecolor{rred}{HTML}{D7191C}

\def\offset{0pt}

\begin{tikzpicture}
  \begin{axis}[
    ybar=.001\linewidth,
    width = 2.8in, 
      height = 1.8in,
    bar width = .02\linewidth,
    symbolic x coords={128, 256, 512, 1024, 2048}, 
    xtick=data,
    legend pos=north west,
    xlabel = Memory~(MB),
    xlabel style={yshift=-1em},
    tick align=outside,
    ylabel style={name=ylabel, yshift=0em},
    ylabel = Completion time (ms),
    enlarge x limits={abs=10pt},
    enlarge y limits=0.1,
    ymax = 400,
    x tick style={xshift=\offset},
    x tick label style={rotate=45, anchor=north east, yshift=\offset},
    cycle list name=exotic,
    legend style={ at={((0.5,1).north)},yshift=1ex, anchor=south, legend columns=-1},
    ]

   \addplot [draw=none,fill=color4,  forget plot, mark = none, error bars/.cd,y dir=plus, y explicit] 
  table[x=name, y expr = \thisrow{mean}, y error expr = \thisrow{max} - \thisrow{mean}, col sep=comma] {figure/data/seclat.csv};
  \addplot [draw=color4,   mark = none,every node near coord/.style={inner ysep=5pt}, error bars/.cd,y dir=minus, y explicit] 
  plot [pattern = dots]
  table[x=name, y expr = \thisrow{mean}, y error expr = \thisrow{mean} - \thisrow{min}, col sep=comma] {figure/data/seclat.csv};

  \addplot [draw=none,fill=rred!80,  forget plot, mark = none, error bars/.cd,y dir=plus, y explicit] 
  table[x=name, y expr = \thisrow{mean}, y error expr = \thisrow{max} - \thisrow{mean}, col sep=comma] {figure/data/gvisorlat.csv};
  \addplot [draw=rred!60,   mark = none,every node near coord/.style={inner ysep=5pt}, error bars/.cd,y dir=minus, y explicit] 
  plot [pattern = north west lines]
  table[x=name, y expr = \thisrow{mean}, y error expr = \thisrow{mean} - \thisrow{min}, col sep=comma] {figure/data/gvisorlat.csv};

\legend{SecLambda, gVisor};
\end{axis}
\end{tikzpicture}

%% file: conclusion.tex
\section{Related Work}\label{sec:related}

Host-based intrusion detection systems~(HIDS) detect potential attacks 
by monitoring an application's execution~\cite{denning1987intrusion}. A specific type of HIDS, model-based HIDS, 
builds a model of the expected execution behavior~(i.e., allowed sequences of system calls) 
of the monitored application, and compares the system calls issued by the application during its execution 
against the model to detect anomalies~\cite{forrest1996sense}. The model is usually represented by an automaton. 
There is a line of research on model-based HIDS, 
focusing on either model construction~(dynamic analysis~\cite{feng2003anomaly}, static code analysis~\cite{wagner2001intrusion}, 
static binary analysis~\cite{giffin2002detecting}, etc.) or model design~(abstract stack model~\cite{wagner2001intrusion}, 
Dyck model~\cite{giffin2004efficient}, inlining model~\cite{gopalakrishna2005efficient}, etc.). 
Control flow integrity~(CFI) can be treated as a special type of intrusion 
detection mechanism for enforcing a nondeterministic finite automaton~(control flow graph) 
to prevent the application from deviating from normal execution paths~\cite{cfi}. 
Our work is inspired by these work and applies model-based intrusion detection and 
CFI to a new setting.  

Information flow control~(IFC) can track and restrict information flow in a system, 
and enforce fine-grained security policies. It has been applied to conventional 
distributed systems~(e.g., DIFC) and cloud applications~\cite{zeldovich2008securing, roy2010airavat, bacon2014information, pasquier2016information, pasquier2017camflow}, so, naturally, 
it can be used for improving serverless security. Trapeze, to the best of our knowledge, 
is the first system to apply 
IFC to serverless applications~\cite{trapeze}. However, Trapeze 
introduces relative high runtime overhead and, more importantly, requires modifying 
all the services being used by the application to support IFC. The requirement of 
homogeneous infrastructure may make it difficult to apply 
Trapeze and other similar IFC mechanisms in real serverless applications, which heavily 
interact with third-party services. Other useful security tools such as event tracing 
provenance actually face the same issue in the serverless 
environment~\cite{xtrace, Orion, pinpoint, pivot, kronos, pip, dapper, magpie, han2018provenance, 
hossain2018dependence, liu2018towards}. Valve, proposed by Datta et al., is 
a secure serverless platform that supports dynamic information flow control~\cite{valve}. 
Compared to Trapeze, Valve provides better performance, application transparency, and usability. 
However, similar to Trapeze, Valve also requires cooperation from third-party services to propagate 
taint labels. Besides, Valve, as well as Trapeze, 
does not consider input-dependent functions during policy generation and flow tracking. 
With regard to performance, Valve intercepts HTTP(S) requests using a MITM proxy, which may introduce more  
overhead compared to directly intercepting system calls as in \sysname. Another 
difference between Valve and our work is \sysname is designed as an extensible security framework that 
can run many different security tasks while Valve focuses on information flow control. 
As future work, we are interested in developing Valve-like security functions in \sysname.   

\section{Conclusion}
In this work, we proposed a new framework, \sysname, for securing serverless applications. 
In \sysname, each function runs in a modified runtime that reports the current 
function state to a guard, which executes 
a set of security functions to perform given security tasks based on the state. 
A centralized controller collects function states of different functions, 
which will serve as input for security functions to facilitate more 
sophisticated security tasks. Using \sysname, we build three security functions 
for modeling and monitoring application behaviors, obfuscating credentials in requests, 
and rate limiting, in order to prevent flow injection attacks, data leakage, and DoS attacks. 
We evaluated the performance of \sysname in a local testbed and its efficiency in AWS Lambda using three 
open-source severless applications, and our results suggest \sysname introduces 
acceptable overhead while providing great security~benefits.